\newcommand{\version}{February 7, 2001}

\documentclass[12pt]{article}
\usepackage{amsmath,amsgen,amstext,amsbsy,amsopn,amsthm,amssymb}

 \swapnumbers
\theoremstyle{plain}
\newtheorem{thm}{THEOREM}

\newtheorem{lem}[thm]{LEMMA}

\theoremstyle{definition}

\newcommand{\infspec}{{\rm inf\ spec\ }}

\newcommand{\R}{{\mathbb R}}

\newcommand{\N}{{\mathbb N}}

\newcommand{\Ll}{{\mathcal L}}
\newcommand{\Hh}{{\mathcal H}}

\newcommand{\x}{{\bf x}}
\newcommand{\y}{{\bf y}}

\newcommand{\xpp}{\x^\perp}
\newcommand{\ypp}{\y^\perp}

\newcommand{\Tr}{{\rm Tr}}
\newcommand{\half}{\mbox{$\frac{1}{2}$}}
\newcommand{\third}{\mbox{$\frac{1}{3}$}}

\newcommand{\rmd}{{d}}
\newcommand{\al}{{\alpha}}

\date{\small\version}

\begin{document}
\markboth{\scriptsize{HS \version}}{\scriptsize{HS \version}}
\title{\bf{Bounds on one-dimensional exchange energies with application
 to lowest Landau band quantum mechanics}}
\author{\vspace{5pt} Christian Hainzl$^1$ and Robert Seiringer$^{2}$\\
\vspace{-4pt}\small{ Institut f\"ur Theoretische Physik,
Universit\"at Wien}\\ \small{Boltzmanngasse 5, A-1090 Vienna,
Austria}}
\date{\small\version}

\maketitle

\begin{abstract}
By means of a generalization of the Fefferman-de la Llave decomposition we
derive a general lower bound on the interaction energy of one-dimensional
quantum systems. We apply this result to a specific class of lowest Landau band wave
functions.
\end{abstract}

\footnotetext[1]{E-Mail: \texttt{hainzl@thp.univie.ac.at}}
\footnotetext[2]{E-Mail: \texttt{rseiring@ap.univie.ac.at}}

\bigskip
An important issue in the quantum mechanics of many interacting
particles is the description of the energy of the system in terms
of the particle density. In particular, a lower bound to the
difference of the interaction energy and its \lq\lq direct\rq\rq\
part is of interest. For the three-dimensional Coulomb potential
it is known that for any $N$-particle wave function $\Psi$ the
so-called Lieb-Oxford inequality \cite{LO81}
\begin{equation}\label{LO}
\langle\Psi|\sum_{i<j}\frac 1{|\x_i-\x_j|} \Psi\rangle\geq
D(\rho_\Psi,\rho_\Psi)-1.68 \int_{\R^3} \rho_\Psi(\x)^{4/3} d^3\x
\end{equation}
holds, where $D(\rho,\rho)=\half\int_{\R^6} \rho(\x)\rho(\y)|\x-\y|^{-1}d^3\x d^3\y$
is the direct part of the energy, and $\rho_\Psi$ is the density
of $\Psi$. This inequality is very useful for many electron
systems, e.g. large atoms, where the \lq\lq exchange term\rq\rq\
$\int\rho_\Psi^{4/3}$ really captures the correct order of magnitude.
However, for states $\Psi$ describing particles in strong magnetic
fields, $\int\rho_\Psi^{4/3}$ is generally much too large, and
(\ref{LO}), although valid for all $\Psi$, is not very useful in
this case.

In the presence of a strong, homogeneous  magnetic field, the
particles are confined to the lowest Landau band, which determines
their motion perpendicular to the magnetic field and makes them
behave essentially like a one-dimensional system. Therefore we
study the one-dimensional analogue of (\ref{LO}) for arbitrary
convex interaction potentials. For short, we shall call the
indirect part of the interaction energy the \lq\lq exchange
energy\rq\rq; it is defined by
\begin{equation}\label{defi}
{\rm Ex}_\psi=\langle\psi|\sum_{i<j}V(|x_i-x_j|)
\psi\rangle-\half \int_{\R^2} \rho_\psi(x)\rho_\psi(y) V(|x-y|)dx
dy .
\end{equation}
In the following, we derive a general lower bound  to ${\rm
Ex}_\psi$. Our method follows closely the proof of the Lieb-Oxford
inequality given in \cite{S95}. The following Lemma replaces the
Fefferman-de la Llave decomposition of the Coulomb potential. We
will then apply this general bound to special potentials of
interest in the study of systems in magnetic fields.

\begin{lem}[Decomposition of $V$]\label{l1}
Let $V:\R_+\to\R$ be twice continuously differentiable,  with
$\lim_{x\to\infty} V(x)=\lim_{x\to\infty}x V'(x)=0$. Then
\begin{equation}\label{dec}
V(x)=2\int_0^\infty V''(2 r) \chi_r * \chi_r (x)dr,
\end{equation}
where $\chi_r(x)=\Theta(r-|x|)$.
\end{lem}

\begin{proof}
A simple computation, using partial integration  and the fact that
$\chi_r * \chi_r (x)=\max\{0,2r-x\}$.
\end{proof}

Although stated only for differentiable  potentials, Lemma
\ref{l1} holds more generally if the derivatives are interpreted
in the sense of distributions. In particular, if $V$ is convex and
tends to zero at infinity, $V''$ defines a Borel measure and
(\ref{dec}) holds. Note also that (\ref{dec}) implies that a
convex $V$ is positive definite.

We now consider $N$-particle wave  functions
$\psi\in\Ll^2(\R^N,d^Nx)$. A vector in $\R^N$ will be denoted by
$(x_1,\dots,x_N)$. Corresponding to $\psi$ its density $\rho_\psi$
is defined as
\begin{equation}
\rho_\psi(x)=\sum_{i=1}^N \int_{\R^{N-1}}
|\psi(x_1,\dots,x_{i-1},x,x_{i+1},\dots,x_N)|^2 dx_1 \dots
dx_{i-1} dx_{i+1} \dots dx_N.
\end{equation}
Note that $\int\rho=N$. Assuming that $\psi$ has finite kinetic
energy, i.e., 
\begin{equation}
\langle\psi|\sum_{i=1}^N-\frac{\partial^2}{\partial
x_i^2}|\psi\rangle<\infty,
\end{equation}
$\sqrt{\rho_\psi}$ is in $\Hh^1(\R)$ by
the Hoffmann-Ostenhof inequality \cite{HH77}, so it is a bounded
and continuous function. In particular, $\rho_\psi\in\Ll^2(\R)$,
so we can define a {\it mean density} $\bar\rho_\psi$ by
\begin{equation}
\bar\rho_\psi=\frac 1N \int_{-\infty}^\infty \rho_\psi(x)^2 dx.
\end{equation}

Given a function $f\in\Ll^p(\R)$, its Hardy-Littlewood maximal
function $f^*$ is defined as 
\begin{equation}\label{defhl}
 f^*(x)=\sup_{r>0}
\frac 1{2r}\int_{x-r}^{x+r} |f(y)| dy. 
\end{equation}
For $p>1$ the inequality
\begin{equation}\label{hlineq}
\|f^*\|_p\leq 2\left(\frac {2p}{p-1}\right)^{1/p} \|f\|_p
\end{equation}
holds for all $f\in\Ll^p(\R)$ \cite{SW71}.

\begin{lem}[General bound on the exchange energy]\label{corr}
Let $V:\R_+\to \R_+$ be a convex function, with
$\lim_{r\to\infty}V(r)=0$. Let $\psi\in \Ll^2(\R^N)$. Then, for
all $\beta(z) \geq 0$,
\begin{equation}\label{corin}
{\rm Ex}_\psi\geq -\half \int_{-\infty}^\infty dz \left( \rho_\psi^{*2}(z)\int_0^{\beta(z)}
V''(r)r^2 dr+\rho_\psi^*(z) \int_{\beta(z)}^\infty V''(r) r dr\right).
\end{equation}
\end{lem}

\begin{proof}
With the aid of Lemma \ref{l1} we can write
\begin{equation}\label{10}
\sum_{i < j} V(|x_i - x_j|) = 2 \int_{-\infty}^\infty dz
\int_0^\infty dr V''(2r) \sum_{i<j} \chi_r(x_i - z) \chi_r (x_j -
z).
\end{equation}
Denote
\begin{equation}\label{11}
\al_{r,z} = \langle \psi | \sum_{i=1}^{N} \chi_r(x_i - z)
\psi\rangle = \int_{z-r}^{z+r} \rho_\psi (x) dx.
\end{equation}
Since $V$ is convex by assumption, we get, for each positive
function $\beta(z)$,
\begin{eqnarray}\nonumber
0&\leq&\langle \psi|\int_{-\infty}^\infty dz \int_{\half\beta
(z)}^\infty dr V''(2r) \left(\sum_{i=1}^N \chi_r(x_i - z)
 - \al_{r,z} \right)^2 \psi \rangle
 \\&\leq& {\rm Ex}_\psi+\int_{-\infty}^\infty dz \int_0^{\half\beta
(z)} dr V''(2r) \al_{r,z}^2 +  \int_{-\infty}^\infty dz
\int_{\half\beta (z)}^\infty dr V''(2r) \al_{r,z},
\end{eqnarray}
where we used the fact that $\chi_r(x)^2 = \chi_r(x)$ and
\begin{equation}
\int_{-\infty}^\infty dz \int_0^\infty dr V''(2r) \al_{r,z}^2 =
\half \int_{\R^2} dx dy V(| x-y|) \rho_\psi (x) \rho_\psi (y)
\end{equation}
by (\ref{dec}) and (\ref{11}). The definition of the maximal
function (\ref{defhl}) implies $\al_{r,z} \leq 2 r \rho_\psi^*(z)$, so we arrive
at (\ref{corin}).
\end{proof}

Of particular interest in the study of particles interacting  with
Coulomb forces in the presence of a strong magnetic field are the
potentials (see e.g. \cite{HS00} and references therein)
\begin{equation}\label{defvmn}
V_{m,n}(z) = \int_{\R^4} \frac {|\phi_m(\xpp)|^2
|\phi_n(\ypp)|^2}{\sqrt{|\xpp-\ypp|^2+z^2}} \rmd^2\xpp \rmd^2\ypp,
\end{equation}
where $\xpp\in\R^2$ and $\phi_m$ denotes the function in the
lowest Landau band with angular momentum $-m\leq 0$, i.e., using polar
coordinates $(r,\varphi)$,
\begin{equation}\label{amf}
\phi_m(\xpp)=\sqrt\frac B{2\pi}\frac 1{\sqrt{m !}}\left(\frac
{Br^2}{2}\right)^{m/2}e^{-i m \varphi}e^{-B r^2/4}.
\end{equation}
Here $B>0$ is the magnetic field strength. We are in particular
interested in the strong field case, where $B\gg \bar\rho_\psi^2$.

Using the relation
\begin{equation}\label{defwm}
B^{-1/2}V_{m,m}(B^{-1/2}z)=\int_0^\infty dq e^{-q |z|} e^{-q^2}
L_m(q^2/2)^2\equiv W_m(z)
\end{equation}
(see \cite{V76}), where $L_m$ are the Laguerre polynomials, one
easily sees that the potentials $V_{m,m}$ are smooth convex
functions away from $z=0$, for all $m\in \N_0$. Hence we can use
Lemma \ref{corr} to get a lower bound on the exchange energy for these
potentials. In the following, we will use the estimate
\begin{equation}\label{w0}
W_m(0)\leq W_0(0)=\sqrt\frac\pi 4<1.
\end{equation}
We do not try to give the best possible constants in the bound stated below,
but concentrate on the asymptotic behavior of ${\rm Ex}_\psi$ for
large $B/\bar\rho_\psi^2$.

\begin{thm}[Exchange energy for $V_{m,m}$]\label{corrthm}
Let $V=V_{m,m}$ be given by (\ref{defvmn}), for some $m\in \N_0$ and
$B>0$. Then, for all $\psi\in\Ll^2(\R^N)$ with $\rho_\psi\in\Ll^2(\R)$,
\begin{equation}\label{corvmn}
{\rm Ex}_\psi\geq - 16 N \bar\rho_\psi \left(
\ln\left[e^3+\frac{B^{1/2}}{\bar\rho_\psi}\right]+2\right).
\end{equation}
\end{thm}

\begin{proof}
{}From (\ref{defvmn}) and (\ref{defwm}) one easily verifies the
estimates
\begin{equation}
W_m''(r)\leq \frac 2{r^2} W_m(r)\quad{\rm and}\quad W_m(r)\leq
\min\left\{\frac 1 r, W_m(0)\right\}.
\end{equation}
Using this and (\ref{w0}) we get
\begin{equation}
\int_\beta^\infty V''(r)r dr \leq \frac 2\beta
\end{equation}
and
\begin{equation}
\int_0^\beta V''(r)r^2 dr \leq 2\left(1+\left[\ln
\beta B^{1/2}\right]_+\right),
\end{equation}
where $[t]_+=\max\{t,0\}$. Choosing
$\beta=\beta(z)=\rho_\psi^*(z)^{-1}$ Lemma \ref{corr} implies that
\begin{equation}
{\rm Ex}_\psi\geq -\int_{-\infty}^\infty
\rho_\psi^*(z)^2\left(2+\left[\ln
B^{1/2}/\rho_\psi^*(z)\right]_+\right) dz.
\end{equation}
Next we use that for $a>0$
\begin{equation}
[\ln a]_+=\inf_{s>0}\frac 1{se} a^s\leq \inf_{0<s<\third}\frac
1{se} a^s.
\end{equation}
Using (\ref{hlineq}) and the fact that $2^{p+1}p/(p-1)\leq 16$ for
$5/3\leq p\leq 2$ this implies
\begin{eqnarray}\nonumber
{\rm Ex}_\psi &\geq& -\int_{-\infty}^\infty
\rho_\psi^*(z)^2\left(2+\inf_{0<s<\third}\frac 1{se}\left(\frac
{B^{1/2}}{\rho_\psi^*(z)}\right)^s\right)dz \\ \nonumber &\geq& -
32\int_{-\infty}^\infty \rho_\psi(z)^2dz - 16\inf_{0<s<\third}
\frac 1{se}
B^{s/2} \int_{-\infty}^\infty \rho_\psi(z)^{2-s} dz \\
\label{inf} &\geq& -16 N \bar\rho_\psi
\left(2+\inf_{0<s<\third}\frac 1{se}\left(
\frac{B^{1/2}}{\bar\rho_\psi}\right)^s\right),
\end{eqnarray}
where we have used that, for $0\leq s\leq 1$, $\int \rho^{2-s}\leq
(\int\rho^2)^{1-s} (\int\rho)^s$ by H\"older's inequality. Now
\begin{equation}
\inf_{0<s<\third}\frac 1{se}a^s\leq \inf_{0<s<\third}\frac
1{se}(e^3+a)^s=\ln(e^3+a),
\end{equation}
which, inserted into (\ref{inf}), proves the Theorem.
\end{proof}

A lower bound on the exchange energy useful for small fields,
i.e., for $B\ll \bar\rho_\psi^2$, can be obtained much easier. A
general inequality, exploiting the positive definiteness of the
potential, gives \cite{T94}
\begin{equation}
{\rm Ex}_\psi\geq -\frac N2 V(0)\geq  -\frac N2 B^{1/2}.
\end{equation}

We remark that the bound on the exchange energy given in Theorem
\ref{corrthm} is indeed nearly optimal for large
$B/\bar\rho_\psi^2$. Namely, if we take for $\psi$ the Slater
determinant of $\varphi_i$,  $1\leq i\leq N$, with
$\varphi_1(x)=(2R)^{-1/2}\Theta(R-|x|)$ and
$\varphi_i(x)=\varphi_1(x+2R(i-1))$ for some $R>0$, the exchange
energy is easily calculated to be
\begin{equation}
{\rm Ex}_\psi=-N\bar\rho_\psi \int_0^{B^{1/2}/
\bar\rho_\psi}W_m(r)\left(1-\frac{\bar\rho_\psi}
{B^{1/2}}r\right)dr,
\end{equation}
which is precisely of the order $N\bar\rho_\psi \ln[
B^{1/2}/\bar\rho_\psi]$ for $B\gg \bar\rho_\psi^2$. The same holds
for the bosonic case, i.e., for $\psi$ the totally symmetric
product of the $\varphi_i$'s.

We now apply our results to a model described by the Hamiltonian
\begin{equation}
H_m=\sum_{i=1}^N\left(-\hbar^2\frac{\partial^2} {\partial x_i^2}-Z
V_m(x_i)\right)+\sum_{i<j} V_{m,m}(x_i-x_j),
\end{equation}
where $V_m$ is defined similarly to (\ref{defvmn}), namely
\begin{equation}\label{defvm}
V_m(z) = \int_{\R^2} \frac {|\phi_m(\xpp)|^2}
{\sqrt{|\xpp|^2+z^2}} \rmd^2\xpp.
\end{equation}
This Hamiltonian is the projection of the full three-dimensional
Hamiltonian for $N$ electrons in the Coulomb field of a nucleus of
charge $Z$ and in a homogeneous magnetic field $B$ onto the space
of functions in  the lowest Landau band with fixed angular
momentum $m$. It acts on the totally antisymmetric functions in
$\Ll^2(\R^N)$. We first estimate the ground state energy
$E_m(N,Z,B,\hbar)=\infspec H_m$. Neglecting the positive interaction
term the one-dimensional Lieb-Thirring inequality (\cite{LT76}; the
constant is taken from \cite{HLW00}) yields, with $\bar
m=\max\{1/4, m\}$,
\begin{eqnarray}\nonumber
E_m(N,Z,B,\hbar)&\geq& -\frac 4{3\pi} \frac 1\hbar Z^{3/2} \int_{\R} V_m^{3/2}\\
&\geq& -\frac {16}{3 \sqrt\pi} \frac 1{\hbar  \bar
m^{1/4}}\frac{\Gamma(5/4)}{\Gamma(3/4)}\frac 1\hbar Z^{3/2}
B^{1/4}
\end{eqnarray}
independently of $N$, where we used that $V_m(x)\leq (\bar
m/B+x^2)^{1/2}$ (\cite{RW00}, Sect. 1.2; the bound for $m=0$
follows easily from Thm. 20 there). However, for $B\gg Z^2$, this
is a very crude estimate, but can be improved using Lemma 2.1 of
\cite{LSY94a} instead. The result is
\begin{equation}
E_m(N,Z,B,\hbar)\geq -\frac {Z^2}{\hbar^2}\left(\left[\sinh^{-1}(\hbar^2 B^{1/2}/ Z \bar
m^{1/2})\right]^2+1+\frac{\pi^2}{12}\right).
\end{equation}
By appropriate variational upper bounds one can show that these
lower bounds indeed capture the correct leading order of the
ground state energy for large $Z$ and $N$.

We are interested in the exchange energy in a state  close to the
ground state of $H_m$. In the following, we will assume that
$\langle\psi|H_m\psi\rangle\leq 0$. The kinetic energy is then
bounded by
\begin{equation}\label{kin}
T_\psi=\langle\psi|\sum_i -\hbar^2\frac{\partial^2}{\partial x_i^2}\psi\rangle\leq
\frac1{1-\lambda}|E_m(N,Z,B,\lambda^{1/2}\hbar)|
\end{equation}
for all $0<\lambda<1$. Moreover, again by the Lieb-Thirring
inequality, $\int\rho_\psi^3\leq 12\pi^{-2}\hbar^{-2} T_\psi$, so
this gives a bound on $\bar\rho_\psi$ by H\"olders inequality,
namely $\bar\rho_\psi\leq
(\int\rho_\psi^3)^{1/2}(\int\rho_\psi)^{-1/2}$. Using the fact
that the bound on the exchange energy given in Theorem
\ref{corrthm} is monotonically increasing in $\bar\rho_\psi$, we
arrive at an explicit bound on ${\rm Ex}_\psi$, which is of the
order
\begin{equation}
{\rm Ex}_\psi\gtrsim -\left\{\begin{array}{l} N^{1/2}Z \left(\frac B{Z^2 \bar
m}\right)^{1/8}\ln\left[N^{4/3}B Z^{-2}\bar m^{1/3}\right] \\
N^{1/2} Z \ln\left[B/ Z^{2}\bar m\right]\ln\left[N B Z^{-2}\right]
\end{array}\right.
\end{equation}
as long as $\langle\psi|H_m\psi\rangle\leq 0$. Note that $B\gg \bar\rho_\psi^2$ is
equivalent to $B\gg Z^{2}N^{-4/3}\bar m^{-1/3}$.

\bigskip
Our method of estimating the exchange energy also applies to the
three-dimensional Coulomb interaction case, if we restrict ourselves to considering
wave functions $\Psi$ that are the total antisymmetrization of
tensor products of functions of the form
\begin{equation}
\psi_m(z_1,\dots,z_{n_m})\prod_{i=1}^{n_m} \phi_m(\xpp_i) ,
\end{equation}
where the $\psi_m$'s are antisymmetric in all variables, with
one-particle density matrix $\gamma_m$, density $\rho_m$, and
$\sum_m n_m=N$. In the following, we will be concerned with the
potentials
\begin{equation}\label{defvbar}
\hat V_{m,n}(z) = \int_{\R^4} \frac
{\phi_m(\xpp)\overline{\phi_n(\xpp)}
\phi_n(\ypp)\overline{\phi_m(\ypp)}}{\sqrt{|\xpp-\ypp|^2+z^2}}
\rmd^2\xpp \rmd^2\ypp.
\end{equation}
Analogously to (\ref{defwm}) there is the relation
\begin{equation}
B^{-1/2}\hat V_{m,n}(B^{-1/2}z)=\int_0^\infty dq e^{-q |z|}
e^{-q^2}\frac{m!}{n!}\left(\half q^2\right)^{n-m}
L_m^{n-m}(q^2/2)^2
\end{equation}
(see \cite{V76}), where the $L_m^n$ are the associated Laguerre
polynomials. From this decomposition we deduce the important
property
\begin{equation}\label{vtilde}
\sum_{n=0}^\infty B^{-1/2}\hat V_{m,n}(B^{-1/2}z)=\int_0^\infty dq
e^{-q |z|} e^{-q^2/2}= B^{-1/2} V_0(B^{-1/2}z) \leq
\min\left\{\sqrt\frac\pi 2,\frac 1{|z|}\right\},
\end{equation}
where $V_0$ is defined in (\ref{defvm}) and the last identity follows again from \cite{V76}.

For $\Psi$ as above, we now estimate the exchange energy. We have
\begin{eqnarray}\nonumber
&&\langle\Psi|\sum_{i<j}\frac 1{|\x_i-\x_j|}\Psi\rangle= \sum_m
\langle\psi_m|\sum_{i<j}^{n_m}V_{m,m}(z_i-z_j)\psi_m\rangle\\ \label{nowest}
&&+\half \sum_{n\neq m}\int_{\R^2}\left(V_{m,n}(z-z')
\rho_m(z)\rho_n(z')-\hat
V_{m,n}(z-z')\overline{\gamma_m(z,z')}\gamma_n(z,z')\right)dzdz'.
\end{eqnarray}
For the last term in (\ref{nowest}) we use (\ref{vtilde}) and 
$|\gamma_m(z,z')|^2\leq \rho_m(z)\rho_m(z')$ to estimate
\begin{eqnarray}\nonumber
&&\sum_{n\neq m}\int_{\R^2}\hat
V_{m,n}(z-z')\overline{\gamma_m(z,z')}\gamma_n(z,z')dzdz'\leq
 \sum_m \int_{\R^2} V_0(z-z')|\gamma_m(z,z')|^2 dz dz' \\ \nonumber&&\leq
\sum_m \int_{|z-z'|\leq \bar\rho_m^{-1}} V_0(z-z')\rho_m(z)\rho_m(z') dz dz'+
\sum_m \int_{|z-z'|\geq \bar\rho_m^{-1}} V_0(z-z')|\gamma_m(z,z')|^2 dz dz'\\ &&
\leq \sum_m n_m \bar\rho_m 
\left(2+\left[\ln\sqrt\frac \pi 2 \frac{B^{1/2}}{\bar\rho_m}\right]_+\right),
\end{eqnarray}
where we used the Cauchy-Schwarz inequality for
the first part the fact that $\int |\gamma_m|^2=\Tr[\gamma_m^2]\leq
\Tr[\gamma_m]=n_m$ for the second.
For the first term in (\ref{nowest}) we apply Thm. \ref{corrthm} (and the fact that 
by an analogous estimate as above this theorem
holds also for $N=1$, where the interaction energy is zero).
Therefore
\begin{equation}
\langle\Psi|\sum_{i<j}\frac 1{|\x_i-\x_j|}\Psi\rangle\geq
D(\rho_{\Psi},\rho_{\Psi}) - \sum_m n_m \bar\rho_m
\left(\frac{33}2
\ln\left[e^3+\frac{B^{1/2}}{\bar\rho_m}\right]+33+\frac 14
\ln[\pi/2] \right).
\end{equation}
Now using  H\"older's inequality for $\bar\rho_m$ and concavity
and monotonicity of $x^{1/2}\ln[e^3+x^{-1/2}]$ in $x$ we conclude
\begin{eqnarray}\nonumber
&&\langle\Psi|\sum_{i<j}\frac 1{|\x_i-\x_j|}\Psi\rangle \geq
D(\rho_\Psi,\rho_\Psi)\\ &&-  N^{1/2} \sqrt{\sum_m
\int_{\R}\rho_m^3}\left(\frac {33}2 \ln\left[e^3+\sqrt{B N/\sum_m
\int_{\R}\rho_m^3} \right]+33+\frac 14 \ln[\pi/2] \right).
\end{eqnarray}
The Lieb-Thirring inequality implies that
\begin{equation}
\sum_m \int_{\R}\rho_m^3\leq \frac
{12}{\pi^2}\sum_m\langle\psi_m|\sum_{j=1}^{n_m}-\frac{\partial^2}{\partial
z_j^2}|\psi_m\rangle =\frac
{12}{\pi^2}\langle\Psi|\sum_{j=1}^{N}-\frac{\partial^2}{\partial
z_j^2}|\Psi\rangle,
\end{equation}
which can be bound as in (\ref{kin}), as long as $\Psi$ has
negative energy. Explicit bounds on the total energy of an atom in a magnetic field 
are given in
\cite{LSY94b}, (2.42) and (2.43), which imply
\begin{equation}\label{hymne}
{\rm Ex}_\Psi\gtrsim -\left\{\begin{array}{ll} Z^{3/5} N^{4/5}
B^{1/5} \ln[B N^{2/3}/Z^{2}] & {\rm for}\quad B\gg Z^2
N^{-2/3}\\ NZ \ln[B/Z^2]\ln[B/Z^2N] & {\rm for}\quad B\gg Z^2 N 
\end{array}\right.
\end{equation}
for this system.

The result in (\ref{hymne}) agrees excellently with the expected
order of magnitude of the exchange energy of wave functions close
to the ground state of large atoms in strong magnetic fields 
(\cite{LSY94a}; see also \cite{HS00}, Remark
6.1). Although our considerations neglect correlations between
particles with different angular momentum, we strongly conjecture
that their contribution is of the same or of lower order, so it
can be expected that (\ref{hymne}) gives the correct order of the
exchange energy in the ground state. 
The bound (\ref{hymne}) in particular applies to all
Slater determinants of angular momentum eigenfunctions in the lowest 
Landau band.

Note that since in the lowest Landau band the density orthogonal
to the magnetic field is bounded by $B/2\pi$, the quantity
$(\sum_m\int\rho_m^3 /N)^{1/2}$ is of the order of $\bar\rho_{3D}
/ B$, where $\bar\rho_{3D}$ is the mean three-dimensional density.
Therefore the result stated above is in total agreement with
investigations on the homogeneous electron gas in a  strong
magnetic field \cite{DG71,FGPY92}, where an exchange energy of the
order $N B^{-1} \bar\rho_{3D} \ln [B^{3/2}/\bar\rho_{3D}]$ was
obtained.

\end{document}